# Elastic properties of bulk and low-dimensional materials using Van der Waals density functional

Kamal Choudhary[1], Gowoon Cheon[2], Evan Reed[2], Francesca Tavazza[1]

1 Materials Science and Engineering Division, National Institute of Standards and Technology, Gaithersburg, Maryland 20899, USA

2 Department of Materials Science and Engineering, Stanford University, Stanford, California 94305, United States

**Abstract:**

In this work, we present a high-throughput first-principles study of elastic properties of bulk and monolayer materials mainly using the vdW-DF-optB88 functional. We discuss the trends on the elastic response with respect to changes in dimensionality. We identify a relation between exfoliation energy and elastic constants for layered materials that can help to guide the search for vdW bonding in materials. We also predicted a few novel materials with auxetic behavior. The uncertainty in structural and elastic properties due to the inclusion of vdW interactions is discussed. We investigated 11,067 bulk and 257 monolayer materials. Lastly, we found that the trends in elastic constants for bulk and their monolayer counterparts can be very different. All the computational results are made publicly available at easy-to-use websites: https://www.ctcms.nist.gov/~knc6/JVASP.html and https://jarvis.nist.gov/ . Our dataset can be used to identify stiff and flexible materials for industrial applications.

**Introduction:**

Mechanical properties describe the response of a material to deformation and are important characteristics in describing solids. From an atomistic perspective, elasticity arises from interatomic bonding and bonding environments. The elastic tensor (ET) [1] is a key property for describing elastic deformation and depends on the symmetry of the material. Important properties such as bulk modulus, shear modulus, Young's modulus, Poisson ratio and sound velocity, universal elastic anisotropy [2] in materials can be easily obtained from the elastic tensor. Further, ET can also be used for determining thermal properties such as heat capacity, Debye temperature, and thermal conductivity [3,4]. Pugh ratio [5] and Pettifor criterion [6,7] obtained from ET can be used to predict ductility and brittleness of materials. Additionally, ET can be used to evaluate the stability of materials in terms of Born's stability criterion [8], elastic energy storage applications [9] and in screening substrates for heterostructure design [10].

Three-dimensional bulk materials (3D), especially those with covalent and metallic bonding environments have been so important in human civilization that ages have been named after them (stone, bronze and iron ages). However, materials in which part of the bonding is due to Van der Waals (vdW) interactions can be considered to reduce their dimensionality, as exfoliation becomes energetically feasible in the vdW direction(s). Therefore, materials with vdW bonding in one, two or three dimensions could be exfoliated down to a two-, one-, or zero-dimensional (2D, 1D, and 0D) counterparts. ET not only varies with materials but can be dependent on materials' dimensionality as well [11]. For example, graphene is the strongest material while graphite is brittle in nature [12,13]. Solids with vdW bonding can exhibit interesting physical properties such as superconductivity [14], charge density waves [15], and the emergence of topological states [16]. In some cases, the physical properties of the material can change as its dimensionality is reduced. For instance, an indirect gap for the bulk system can become direct in the monolayer case. Similarly, it is not unreasonable to assume that elastic property may show similar trends depending on the bulk vs monolayer materials. However, comparison of bulk and monolayer elastic constants is not trivial, as their units change from Pa (or $Jm^{-3}$) for the bulk case to units of $Jm^{-2}$ (or $Nm^{-1}$) for the monolayer case. Also, the elastic response becomes more complex for monolayer materials, as it may become thickness dependent such as for $MoS_2$ [17,18]. High demand for flexible and miniaturized electronics requires a thorough insight into both bulk and monolayer elastic properties, but it is difficult to obtain such information by experiments only. While experiments [19,20] such as ultrasonic measurement and nanoindentation can be used to measure the ET for bulk and low dimensional materials, their scope is limited to only a small number of available experimental data. A possible solution to this experimental limitation is to use computationally reliable tools such as density functional theory (DFT) to calculate ET, as they can be applied to thousands [21] of compounds in a reliable way and in a realistic time-frame. In fact, exotic phenomenon such as negative Poisson ratio for two-dimensional black phosphorous was first predicted by density functional theory [22] and only later verified by experiments [23].

In the literature, there are only a few systematic studies of dimension dependent ET such as the works of Duerloo et al. [24] and Gomes et al. [25] but a large database of monolayer materials is still needed. While much work has been done towards building consistent DFT databases for bulk materials' ET, as, for instance, the VLab project [26] and the Materials Project (MP) [21], however, these datasets do not contain dimension dependent elastic properties such as mono and multi-layer ETs. Additionally, these datasets use homogeneously fixed DFT plane wave parameters (plane wave cut-off and number of k-points for sampling the Brillouin zone), which is not necessarily the best computational choice to get high accuracy evaluations of ET, especially for vdW bonded materials [27]. Most importantly, generalized gradient based exchange-correlation functional (such as Perdew-Burke-Ernzerhof, PBE) is generally used in these databases, which is not suitable for vdW bonded materials [28,29]. Recently, the lattice constant error criteria [30] , data-mining approaches [31], topological scaling algorithm [32] and geometric and bonding criteria [33] have been used to demonstrate that around 5000 materials are vdW bonded, which implies that there is a real necessity to evaluate their elastic properties using suitable DFT methodologies. Moreover, it is important to evaluate the performance of vdW functionals such as vdW-DF-optB88/OptB88vdW (OPT) [28,29,34-36] for non-vdW materials compared to PBE in a systematic way.

In this work, we addressed these issues by calculating, the elastic constants of 11,067 bulk and 257 monolayer materials using a vdW functional (OPT) and material-dependent cutoff and k-point (DFT parameters) to guarantee a controlled level of convergence in all cases. Our results are posted on the JARVIS-DFT website (https://www.ctcms.nist.gov/~knc6/JVASP.html ). The REST-API [37] is available at https://jarvis.nist.gov . Due to our high-throughput approach, we have sufficient data to meaningfully investigate trends in elastic constants-derived properties, such as bulk and shear modulus, Poisson ratio and Pugh ratio. Additionally, we investigate the vdW bonding (in terms of exfoliation energy) relation with elastic constants of layered 2D-bulk materials.

The paper is organized as follows: first we present the methodology used in our DFT calculations, then we discuss our results for bulk materials that are predicted to be vdW bonded in three dimensions (referred to as "0D" material in the rest of the paper), in two dimensions ("1D" materials), in one dimension ("2D" materials) and no vdW bonding at all (referred to as "3D" or bulk materials in the rest of the paper). It is emphasized that dimensionality is interpreted mainly to distinguish whether the materials have vdW bonding or not. Unless specified as monolayer (1L), the materials are periodic in three dimensions during DFT calculations. Monolayer materials are non-periodic in *z*-direction. Following discussion of three-dimensional periodic materials, we describe elastic constants for monolayer (1L) materials. We also investigate the ET relation of monolayers and their bulk counterparts.

## Method:

The DFT calculations are performed using the Vienna Ab-initio Simulation Package (VASP) [38,39] and the projector-augmented wave (PAW) method [40]. Please note that commercial software is identified to specify procedures. Such identification does not imply recommendation by the National Institute of Standards and Technology. The crystal structures were mainly obtained from Materials Project (MP) DFT database [21]. More specifically, we obtained all the crystal structures obtained for the optoelectronic database [41], potential candidates for layered materials that we identified with lattice-constant approach [30] and data-mining approach [31]. The data mining-approach is based on the difference in bond-lengths in vdW bonded solids compared to other non-vdW bonded materials. The data-mining approaches also identified several mixed-dimensional materials. The lattice constant criterion is based on the difference in lattice constant prediction between DFT and experimental data. Specifically, the large difference in lattice constant (compared to experiment or suitable vdW functional) is encountered if non-vdW-including functional (such as PBE) is used for simulating vdW bonded solids, such as $MoS_2$. So, the lattice constant criteria predict that if there is a large difference in lattice constant prediction, then it should be vdW bonded (for non-cubic systems). If the difference is large (5% or more) in only one lattice direction, the material could be 2D-bulk, if the difference is large in two directions then it could be 1D-bulk and if there is a large difference in lattice constants for all three directions, then it could be 0D-bulk material. The 2D-bulk materials can be exfoliated in one direction (with vdW bonding) to form 2D-monolayer/multilayer. The 1D-bulk can be exfoliated in two directions for 1D-molecular chain. Similarly, the 0D-bulk can be exfoliated in three directions to a quantum dot-like material. Examples of dimensionality in materials, as discussed above, is shown in Fig. 1. The exfoliation is feasible due to the weak vdW bonding [42]. In the previous work [30], this simple criterion was shown successful to 89% accuracy by actual exfoliation energy calculations.

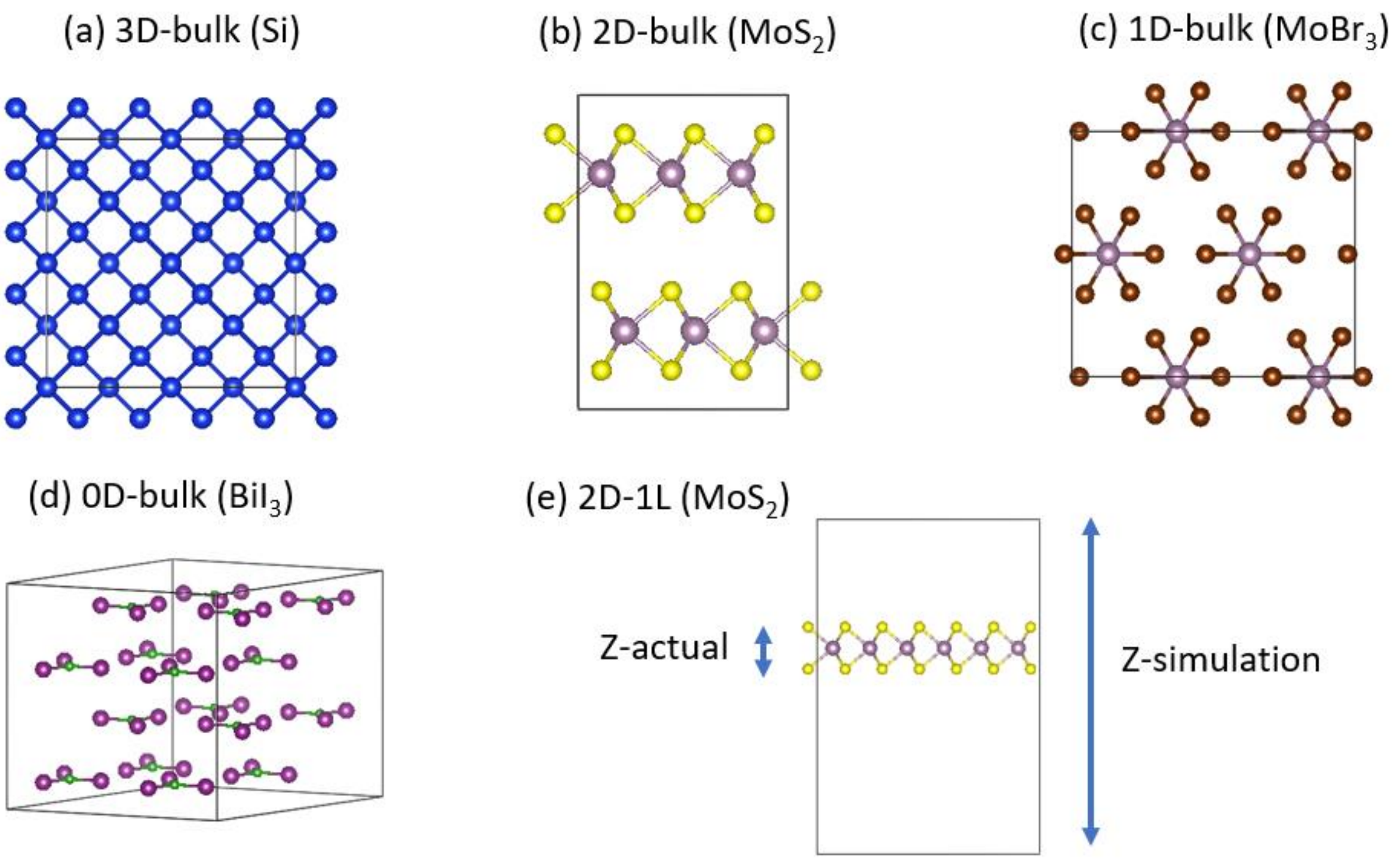

*Fig. 1 Figure showing different classes of materials. Examples for a) 3D-bulk diamond Si, b) 2D-bulk 2H-$MoS_2$, c) 1D-bulk $MoBr_3$, d) 0D-bulk $BiI_3$ and e) 2D-1L ($MoS_2$ monolayer) are shown. Dimensionality is reduced due to the presence of vdW bonding in one, two or three crystallographic dimensions.*

Next, it is important to select a DFT functional which can describe both vdW bonded and non-vdW bonded materials with reasonable accuracy. The dispersion or van der Waals interactions are due to electronic density fluctuations of distant regions in space. The dispersion force, which originates from the nonlocal electron correlation, can be described by using post-Hartree-Fock quantum chemistry methods, such as Møller-Plesset perturbation theory [43]; coupled cluster with singlet, doublet, and perturbative triplet [CCSD(T)] [44]; quantum Monte Carlo [45]; and the adiabatic-connection fluctuation dissipation theorem (ACFDT) [46,47]. However, solving the Hamiltonian for the above methods corresponds to solving the full many-body problem, and is unfeasible for realistic systems, unless an approximation to the exchange-correlation kernel is found. Recently, there has been an increasing interest in adding van der Waals correction to DFT [48,49] . A wide variety of new types of methods have been developed and applied successfully to a broad range of systems. Some of them include DFT+D [50], Tkatchenko-Scheffler (TS) methods [51], vdW-DF methods [34,52-59], Vydrov and Van Voorhis (VV10) method [60]. The vdW-DF is a promising approach, as it depends only on the charge density n(r) and its gradient |∇n(r)| without empirical fitting parameters like DFT+D. In addition, it is able to describe the dispersion (or van der Waals (vdW)) forces and covalent bonding in a seamless way. The exchange-correlation energy within vdW-DF is given by:

$$E_{xc} = E_x^{GGA} + E_c^{LDA} + E_c^{NL} \tag{1}$$

where $E_x^{GGA}$ is the exchange energy within the generalized gradient approximation (GGA) and $E_c^{LDA}$ is the correlation energy within the local-density approximation (LDA). The nonlocal correlation energy is given by:

$$E_c^{NL} = \frac{1}{2}\iint d\boldsymbol{r}d\boldsymbol{r}'n(\boldsymbol{r})\phi(d,d')n(\boldsymbol{r}') \tag{2}$$

where $\phi$ is a kernel function, $d = q_0(r)|r - r'|$ and $d' = q_0(r')|r - r'|$. The $q_0$ is a function of n(r) and |∇n(r)|, and it is proportional to the gradient corrected LDA exchange-correlation energy per electron. This function controls the behavior of $E_c^{NL}$ in the slowly varying as well as nonuniform density regions. It is noted that the use of the LDA correlation is motivated by the fact that $E_c^{NL}$vanishes in the uniform electron gas limit, and to avoid the possible double counting of the gradient correction contained in $E_c^{NL}$. Hence, the vdW-DF-optB88 is an example of the truly nonlocal-correlation functionals in the vdW-DF-method for approximating the vdW forces in regular DFT.

In this work, we use vdW-DF-optB88/OptB88vdW (OPT) functional for structure, energetics and elastic property calculations. The OPT exchange functional uses the Becke88 (B88) exchange [61] and optimizes the parameters in the B88 enhancement factor. The OPT functional was shown to be very well applicable to solids in ref. [29] and, ever since, it has been used to model rare-gas dimers and metallic, ionic and covalent bonded solids [29,49], polymers [62] and small molecular systems [63]. As we obtained the crystal structures from MP, which uses PBE functional, we re-optimized those structures with OPT because the error in lattice constants can significantly influence the error in the calculation of elastic properties [28,29].

We performed plane-wave energy cut-off and k-point convergences with 0.001 eV tolerance on energy for each structure in an automated way. The structure relaxation with OPT functional was obtained with $10^{-8}$ eV energy tolerance and 0.001 eV/Å force-convergence criteria. During elastic constants calculations, we further increase the plane-wave cut-off by 30 %. The elastic tensor is determined by performing six finite distortions of the lattice and deriving the elastic constants from the strain-stress relationship [64,65]. A set of strains $\boldsymbol{\varepsilon}=(\varepsilon_1,\varepsilon_2,\varepsilon_3,\varepsilon_4,\varepsilon_5,\varepsilon_6)$ where $\varepsilon_1$, $\varepsilon_2$, and $\varepsilon_3$ are the normal strains and the others are the shear strains imposed on a crystal with lattice vectors **R** specified in Cartesian coordinates,

$$\mathbf{R}=\begin{pmatrix} a_1 & a_2 & a_3 \\ b_1 & b_2 & b_3 \\ c_1 & c_2 & c_3 \end{pmatrix} \tag{3}$$

where $a_1$ is the x-component of the lattice vector $\vec{a}$, $b_2$ the y-component of the lattice vector $\vec{b}$ , and so on. Corresponding to a set of strains discussed above, a set of stresses **σ** $=(\sigma_1, \sigma_2, \sigma_3, \sigma_4, \sigma_5, \sigma_6)$ are determined with VASP code. The stress-strain can then be related by general Hooke's law:

$$\boldsymbol{\sigma} = \mathbf{C}\boldsymbol{\varepsilon} \tag{4}$$

where **C** is a 6x6 elastic constant matrix [66], which can be obtained by matrix-inverse operations.

ET is determined with spin-unpolarized ET calculations except for materials containing magnetic elements for which brute-force spin-polarized calculations are required for reasonable ET data (especially for Fe and Mn compounds). We use conventional cells of systems during ET calculations. For bulk material, the compliance tensor can be obtained by:

$$s_{ij} = C_{ij}^{-1} \tag{5}$$

Now, several other elastic properties calculated from $C_{ij}$ and $s_{ij}$. Some of the important properties are given below:

$$K_V = ((C_{11}+C_{22}+C_{33}) + 2(C_{12}+C_{23}+C_{31}))/9 \tag{6}$$

$$G_V = ((C_{11}+C_{22}+C_{33}) - (C_{12}+C_{23} + C_{31}) +3 (C_{44}+C_{55}+C_{66}))/15 \tag{7}$$

$$K_R = ((s_{11}+s_{22}+s_{33}) + 2(s_{12}+s_{23}+s_{31}))^{-1} \tag{8}$$

$$G_R = 15(4(s_{11}+s_{22}+s_{33}) - 4(s_{12}+s_{23}+s_{31}) + 3(s_{44}+s_{55}+s_{66}))^{-1} \tag{9}$$

$$K_{VRH} =(K_V+K_R)/2 \tag{10}$$

$$G_{VRH} =(G_V+G_R)/2 \tag{11}$$

$$\nu = (3K_{VRH} - 2G_{VRH})/(6K_{VRH}+2G_{VRH}) \tag{12}$$

Here $K_V$ and $G_V$ are Voigt bulk and shear modulus, and $K_R$ and $G_R$ Reuss-bulk and shear modulus respectively. The homogenous Poisson ratio [21] is calculated as ν. The EC data can be also used to predict the ductile and brittle nature of materials with Pugh [5] (Gv/Kv) and Pettifor criteria ($C_{12}$-$C_{44}$) [6,7]. Materials with Pugh ratio value >0.571 and Pettifor criteria <0 should be brittle, while materials with Pugh ratio value <0.571 and Pettifor criteria >0 should be ductile [7].

For monolayer material calculations, the elastic tensor obtained from DFT code such as VASP, assumes periodic-boundary-condition (PBC). Therefore, cell vectors are used to calculate the area which again is used in computing stress. When dealing with the monolayer, an arbitrary vacuum padding is added in one of the direction (say z-direction). When computing EC we need to correct the output by eliminating the arbitrariness of the vacuum padding. We do that as a post-processing step by multiplying the $C_{ij}$ components $(i, j \neq 3)$ by the length of the vacuum padding. Therefore, the units of EC turn into $Nm^{-1}$ from $Nm^{-2}$. For example, in order to calculate $C_{11}$ (stress computed in x direction), the area is computed using normal of y and z-vectors. Obviously, the z-vector is arbitrary, so if we multiply the output by z-vector magnitude we get rid of the arbitrariness of z and also get $C_{11}$ in $Nm^{-1}$. As shown in Fig. 1, the z-vector magnitude is the z-simulation. The above discussion can also be expressed as the following:

$$\sigma_{VASP} = \frac{F}{A} = \left.\frac{F}{|z||l|}\right|_{l\in(x,y)} \quad (13)$$

$$\sigma_{mono} = |z| \times \sigma_{VASP} \quad (14)$$

**Results and discussions:**

As discussed in the method section, the crystal structures were obtained from MP, which uses PBE for structure optimization. After convergence of DFT parameters (plane wave-cut-off and k-points), we re-optimize the MP crystal structures with OPT functional. Most of the MP crystal-structures have Inorganic Crystal Structure Database (ICSD) identifiers (IDs), which can be used to obtain experimental lattice parameter information. Hence, we compute PBE and OPT based mean absolute error (MAE) and root-mean-square error (RMSE) compared to experimental data from ICSD in lattice constants of all the available structures in our database. There are presently 10,052 structures with ICSD IDs in our database. We further classify these structures into predicted vdW and predicted non-vdW structures. We use the lattice-constant criteria [30] and data-mining approaches [31] to identify vdW structures. All the remaining structures are treated as non-vdW bonded. The predicted vdW bonded materials can have vdW bonding in one, two or three crystallographic directions. It is to be noted that exfoliation energy is calculated to predict vdW bonded materials [30], but the two heuristic methods mentioned above can act as pre-screening criteria for determining vdW bonded structures. Out of 10,052 structures, 2,241 were predicted to be vdW bonded. We calculate the MAE and RMSE for all the materials, vdW bonded and non-vdW bonded materials as shown in Table 1. As evident from the Table. 1, the OPT seems to improve lattice constants in *a*, *b*, *c* crystallographic directions compared to PBE. Significant improvement in lattice parameters is observed for predicted vdW materials, especially in *c*-directions. For predicted non-vdW materials, the errors are similar for OPT and PBE, suggesting that OPT an improved lattice constant predictions for vdW materials without much affecting the predictions for non-vdW bonded materials. Similar MAE values were obtained for PBE by Tao et al. [67] suggesting agreement in uncertainty-trends.

*Table 1. Mean absolute error (MAE, Å) and root-mean-square error (RMSE, Å) in a, b and c crystallographic directions computed for all materials in our database with respect to experimental data (ICSD data). To facilitate comparison between the functionals, both MAE and RMSE have been computed for all materials, only for predicted vdW bonded materials and only for predicted non-vdW bonded materials, using Material's project PBE and JARVIS-DFT OPT functional.*

| | **#Mats.** | **MAE (*a*)** | **MAE (*b*)** | **MAE (*c*)** | **RMSE (*a*)** | **RMSE (*b*)** | **RMSE (*c*)** |
|---|---|---|---|---|---|---|---|
| **OPT (All)** | 10052 | 0.11 | 0.11 | 0.18 | 0.29 | 0.30 | 0.58 |
| **PBE (All)** | 10052 | 0.13 | 0.14 | 0.23 | 0.30 | 0.29 | 0.61 |
| **OPT (vdW)** | 2241 | 0.20 | 0.21 | 0.44 | 0.44 | 0.44 | 0.99 |
| **PBE (vdW)** | 2241 | 0.26 | 0.29 | 0.62 | 0.45 | 0.51 | 1.09 |
| **OPT (non-vdW)** | 7811 | 0.08 | 0.08 | 0.11 | 0.23 | 0.24 | 0.39 |
| **PBE (non-vdW)** | 7811 | 0.09 | 0.09 | 0.12 | 0.22 | 0.25 | 0.36 |

At present, we have computed elastic constants for 11,067 bulk materials (containing 3D-bulk, 2D-bulk, 1D-bulk and 0D-bulk materials) and 257 monolayers in our database, and the database is still increasing. In Fig. 2a we show the distribution of crystal structures for which the elastic constants were calculated. We observe that cubic and tetragonal structures mainly dominate the database. The other major structure types are orthorhombic and hexagonal, while triclinic crystal system materials are less prevalent. The investigated materials can also be classified according to their predicted dimensionality. The dimensionality prediction of materials is based on the results from the data-mining and the lattice-constant criteria discussed above. These results are displayed in Fig. 2b. Exfoliation energy calculation is computationally the final step to confirm the vdW bonding strength of these predicted materials, and previous results [30], where such calculations were carried out for 430 materials, indicated a ~90 % accuracy for the lattice-constant criteria. Among the investigated materials, 17.4 % are predicted to be vdW bonded: 11.85 % are 2D-bulk, while 1D and 0D materials are only 2.31% and 3.25 %, respectively. Please note that these percentage distributions were determined based on the number of completed elastic constant calculations in our database.  All the materials from the lattice constant criteria, data mining approach and screening of optoelectronic materials are subjected to DFT calculations, and as the calculations get completed (dependent on their cell size, number of electrons etc.) they will be updated on the website.  From the above results, we clearly see the need of calculation of ET with suitable vdW functions such as OPT.

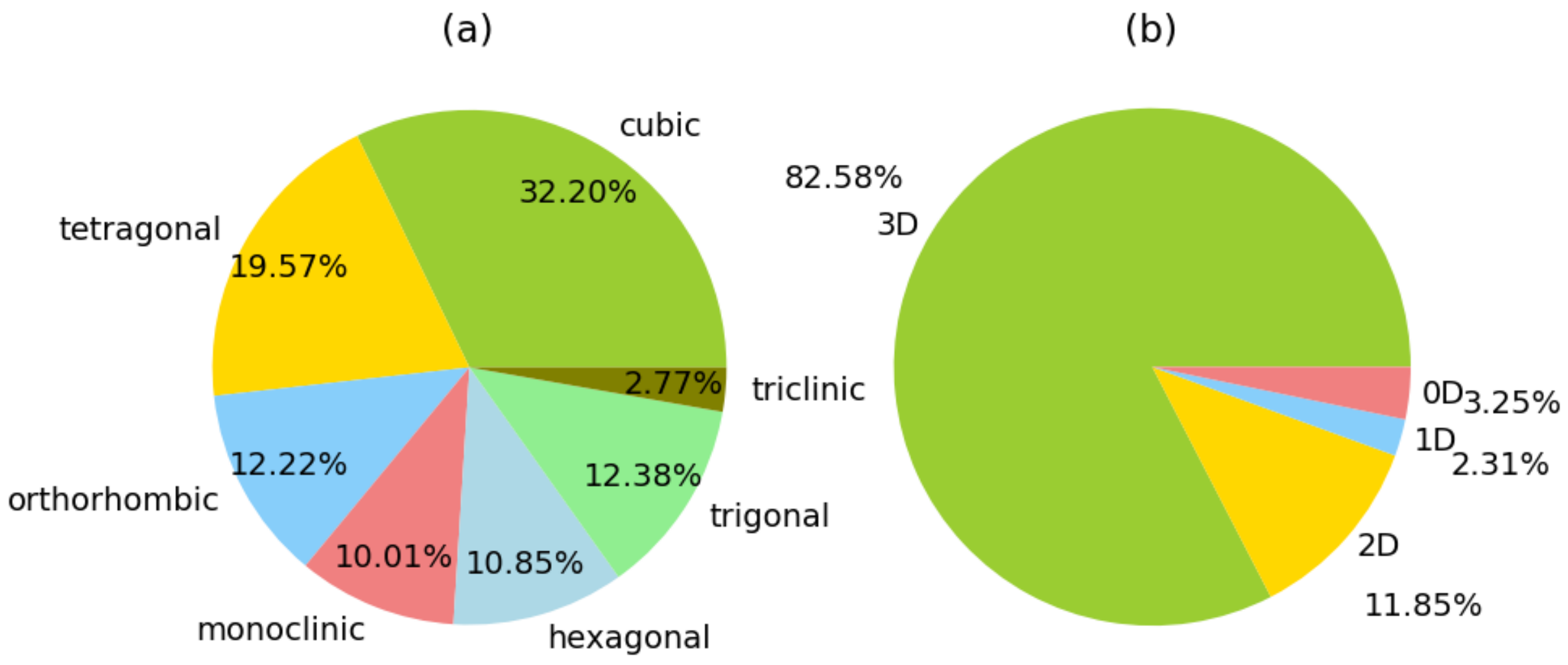

*Fig. 2 a) Crystal-system and b) dimensionality distribution for materials in our database.*

The next step, however, is to investigate whether OPT is reliable in predicting ET properties for general solids. Hence, we compare our bulk modulus data with the experiment in Table. 2. The overall mean absolute error for bulk modulus using the data in the Table. 2 was found as 8.50 GPa.

The experimental data are however not corrected for zero-point energy effects, which would lead to a slight increase of their values [29,68] as the DFT data are computed at 0 K. In order to investigate the effect of neglecting the temperature dependence of elastic constants, we compared DFT $C_{11}$ data to low-temperature experimental data as well as room-temperature data [69] (Table. S2). We find that the mean absolute error in $C_{11}$ ranges from 7.97 to 10.9 GPa for OPT, depending on the temperature of the experimental data of comparison. This indicates that the thermos-physical effects in EC are small and that, overall, the OPT functional can predict bulk modulus of ionic, covalent and vdW bonded materials well. To understand the effect of different flavors of vdW-DF [34,52-59] method, we compared bulk modulus of several materials with several functionals: vdW-DF-optB88 (OPT) [70], vdW-DF-optB86b (MK) [29], vdW-DF-optPBE (OR) [70] and vdW-DF-cx13 (CX) [52]. We find that the vdW-DF functionals give very similar MAEs [69] (Table S1). We also compare properties for a small set of materials with experiment, and these results are provided in the supplementary information [69] (Table S3). The mean absolute error in individual elastic constants could be upto 15 GPa.

*Table. 2 Comparison of bulk modulus, $K_V$ (GPa), from vdW-DF-optB88 (OPT) and experiments. The experimental data are however not data corrected for zero-point energy effects, which would lead to a slight increase of the values [29,68]. The experimental data is taken from refs. [29,71,72].*

| **Material** | **JVASP#** | **OPT** | **Expt.** | **Material** | **JVASP#** | **OPT** | **Expt.** |
|---|---|---|---|---|---|---|---|
| **Cu** | 14648 | 141.4 | 142 | **V** | 1041 | 183.4 | 161.9 |
| **C (diamond)** | 91 | 437.4 | 443 | **Fe** | 882 | 193 | 168.3 |
| **Si** | 1002 | 87.3 | 99.2 | **Ni** | 14630 | 200.4 | 186 |
| **Ge** | 890 | 58.1 | 75.8 | **Nb** | 934 | 176 | 170.2 |
| **Ag** | 813 | 100.3 | 109 | **Mo** | 925 | 262 | 272.5 |
| **Pd** | 14644 | 176 | 195 | **Ta** | 14750 | 199 | 200 |
| **Rh** | 14817 | 260.8 | 269 | **W** | 14830 | 305.2 | 323.2 |
| **Li** | 913 | 13.9 | 13.3 | **Ir** | 901 | 348 | 355 |
| **Na** | 25140 | 7.7 | 7.5 | **Pt** | 972 | 251.6 | 278.3 |
| **K** | 14800 | 3.9 | 3.7 | **Au** | 825 | 148 | 173.2 |
| **Rb** | 978 | 3.1 | 2.9 | **Pb** | 961 | 42.6 | 46.8 |
| **Ca** | 846 | 17.7 | 18.4 | **LiCl** | 23864 | 35.5 | 35.4 |
| **Sr** | 21208 | 12.5 | 12.4 | **NaCl** | 23862 | 27.7 | 26.6 |
| **Ba** | 831 | 9.9 | 9.3 | **NaF** | 20326 | 53.7 | 51.4 |
| **Al** | 816 | 70 | 79.4 | **MgO** | 116 | 160.7 | 165 |
| **LiF** | 1130 | 73.9 | 69.8 | **SiC** | 182 | 213.3 | 225 |
| **$TiO_2$-anatase** | 314 | 196 | 191.9 | **GaAs** | 1174 | 62 | 75.6 |
| **$TiO_2$-rutile** | 10036 | 226.3 | 243.5 | **P (black)** | 7818 | 41 | 36 |
| MAE (GPa): | 8.51 | | | | | | |

Next, we compare in Fig. 3 bulk modulus and shear modulus obtained using OPT to PBE results from the MP database, for all materials common to both databases. We find that the OPT results have an overall excellent agreement with MP data, with Pearson coefficient up to 0.95. The Pearson correlation coefficient (PC) is used to measure the linear correlation between two variables/datasets. It acquires a value between +1 and −1, where 1 is total positive linear correlation, 0 is no linear correlation, and −1 is total negative linear correlation. A PC-value of 0.95 implies that the OPT functional can be used for studying ET for vdW as well as non-vdW bonded materials. In Fig. 3a and Fig. 3b we also show ±15 % deviation from MP, and we find that most of the JARVIS-DFT and MP are within 15 % of each other. To investigate if there is a systematic difference in predictions for low-dimensional properties, we color code the JARVIS-DFT data for predicted low dimensional materials as red dots while the others are depicted in green dots. We observe that both bulk and shear modulus are underestimated using PBE data for predicted low-dimensional materials, with respect to OPT results. The MP data is depicted as a straight line in both the plots. For a perfect agreement between JARVIS-DFT and MP, the green and red dots would lie exactly on the $x = y$ straight line. Hence, we demonstrate that OPT provides a very accurate prediction for both vdW and non-vdW bonded materials. Interestingly, the shear modulus deviates more than bulk modulus data for OPT vs PBE as seen in Fig. 3b. This is mainly because it is generally difficult to obtain the shear properties for vdW materials if vdW interaction is not included. We investigate the materials which were underestimated in OPT. Some of them are: VOF (JVASP-30457), body-centered Si (JVASP-25064), MoTi (JVASP-37029) and O (JVASP-25109). The differences can be attributed to the difference in k-points and plane wave cut-off between OPT and PBE calculations. Our database successfully reproduces some of the widely known high bulk modulus materials, such as $C_3N_4$ [73,74] (445 GPa, JVASP-9141) and diamond (438 GPa, JVASP-25274). Some of the other high bulk modulus materials are: Os (395 GPa, JVASP-14744), OsC (383 GPa, JVASP-15755), $BC_2N$ (379 GPa, JVASP-8703), BN (378 GPa, JVASP-7836), WN (377 GPa, JVASP-19932), Re (364 GPa, JVASP-981), OsN (363 GPa, JVASP-14094), MoN (354 GPa, JVASP-16897), $WIr_3$ (353 GPa, JVASP-18731), MoC (350 GPa, JVASP-14490), Ir (348 GPa, JVASP-901), $IrN_2$ (348 GPa, JVASP-9153), $CoRe_3$ (340 GPa, JVASP-11984), $MoIr_3$ (340 GPa, JVASP-16565), BW (339 GPa, JVASP-14930), $Re_3Ni$ (331 GPa, JVASP-11982).

While it has been established in the literature that the shear modulus could be roughly related to the bonding nature of the materials (for instance metals have lower shear modulus than covalent materials) [75] in case of vdW bonded materials there is no such a clear trend. Interestingly, for low dimensional materials, the shear modulus can attain both very high (such as graphite, 220 GPa, JVASP-48) or very low (such as $P_4S_3$, 6 GPa, JVASP-4346) values. We find that the maximum bulk and shear modulus values are 70 GPa and 24 GPa (JVASP-32164) for 0D, 124 GPa and 101 GPa (JVASP-21473) for 1D, 281 GPa and 220 GPa (JVASP-48) for 2D indicating that the elastic moduli increase as the dimensionality of the bulk material increases. We discuss the effect of dimensionality on elastic properties in more details in the later section.

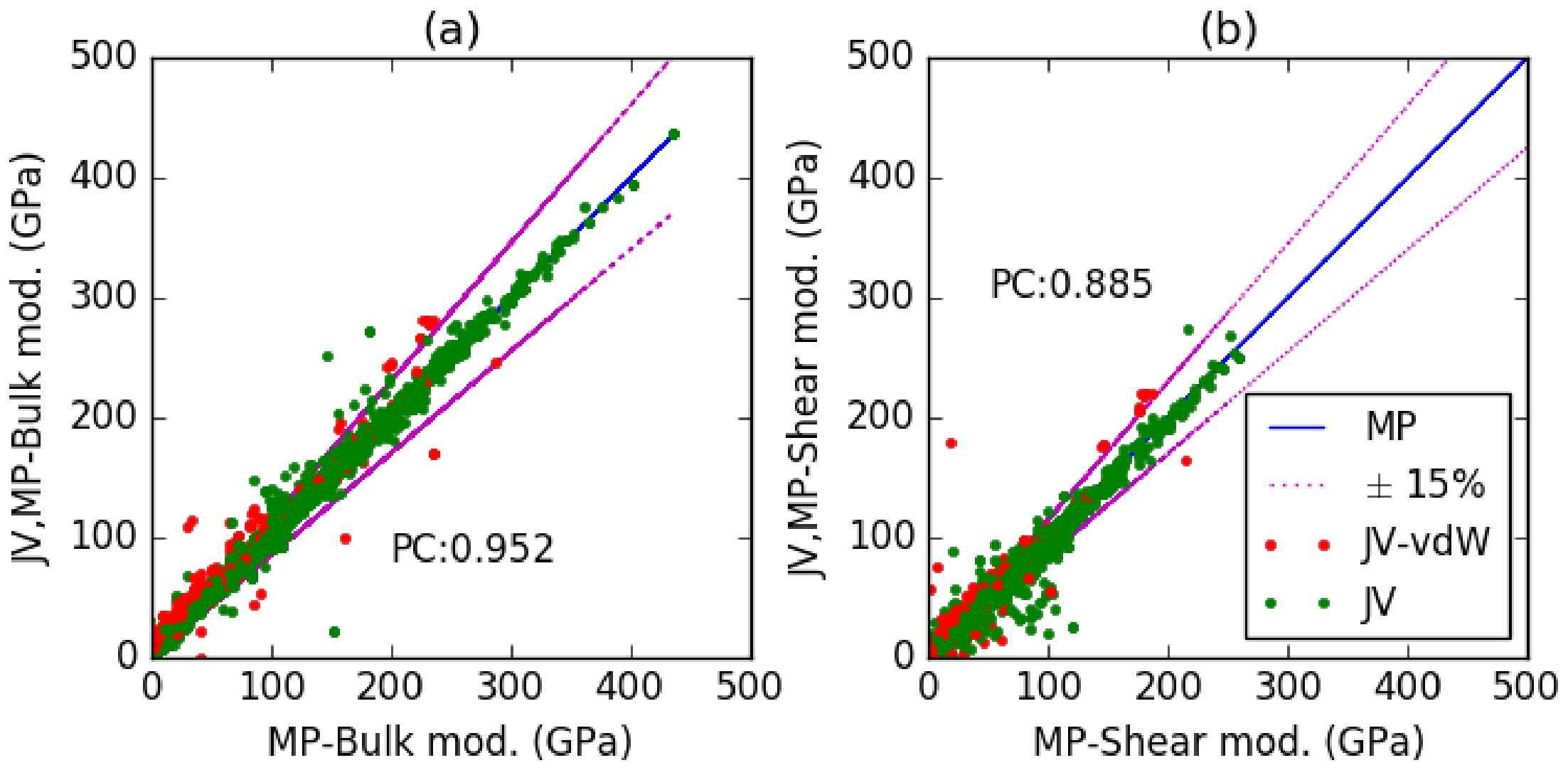

*Fig. 3 Comparison of Voigt (a) bulk and (b) shear modulus obtained from JARVIS-DFT(JV) OPT and Materials project (MP) PBE data. The red dots are moduli for predicted low-dimensional bulk materials, while green dots are for the remaining materials, i.e. the non vdW-bonded materials. Pearson coefficient close to unity suggests excellent agreement in the two datasets.*

Next, we investigated which elements from the periodic table mainly contribute to high bulk modulus materials. We projected the bulk modulus of elements as well as binary, ternary etc. compounds on the individual constituent elements and calculated their average for each element in the periodic table. The trends in the periodic table are shown in Fig. 4. Some of the common high bulk modulus contributing elements found are Re, Os, Ir, B, C, N, O, Tc, Rh, and Ru. This agrees with commonly known high bulk modulus materials as discussed previously. Similar trends were found for the shear modulus data (in Fig. S1, see supplementary information [69]). The high modulus trend for contributing elements near Re and Os in the Fig. 4 can be explained based on the number of half-filled valence *d*-orbitals (as shown in Fig. 5). Similar trends have been observed in the literature for transition metal-nitrides and carbides [76,77]. The periodic table trend results found here can be used as an initial guideline for designing high-strength materials.

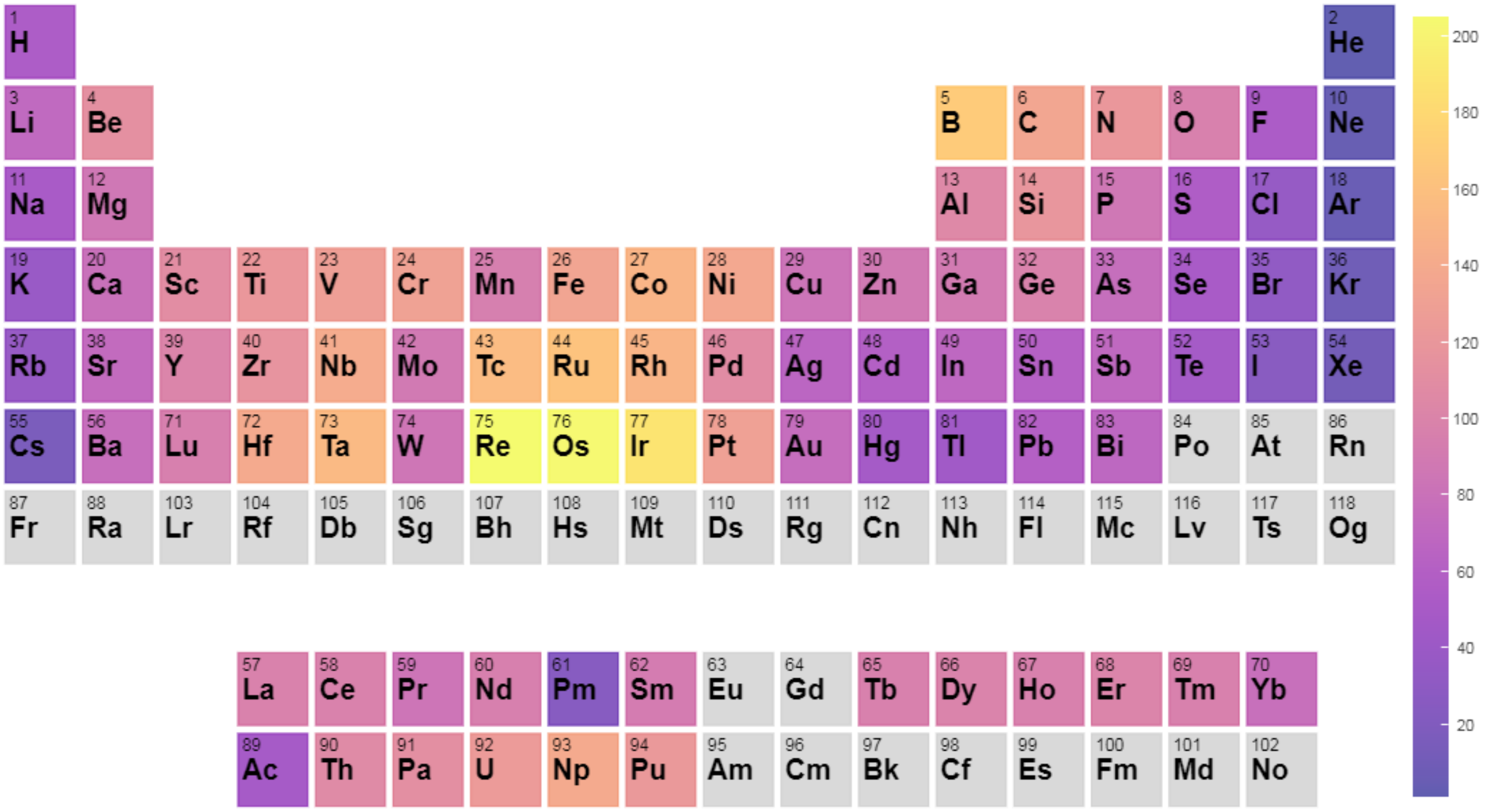

*Fig. 4 Periodic table trend for high bulk modulus material constituents. The bulk moduli of all the materials were projected on individual elements and their average contribution is shown. The colorbar is in the unit of GPa. A similar trend was found for shear modulus.*

Now, we correlate the number of filled d-orbitals with the bulk modulus obtained by averaging the element projected bulk modulus for transition metals (shown in Fig. 4) over each periodic table column. We find that as the number of filled d-orbitals increases, the average bulk modulus increases upto d=6 and then it decreases. The trend found here is consistent with the work in Refs. [76,77] for carbides and nitrides. This is interesting because we didn't just study carbides and nitrides, but all classes of materials together. However, there is a drop in Fig. 5 for d=4 (Cr, Mo and W). We interpret that this drop is due to the over-sampling of vdW-bonded materials containing Mo and W in our database. As vdW bonded materials have low bulk-modulus (as discussed above), over-sampling them would correspond to an unphysical drop in average bulk modulus. As we calculated the percentage of vdW bonded materials containing either Cr, Mo or W in our database, we found it to be 12 %, 62% and 66 % for Cr, Mo and W respectively, indicating an over-sampling of vdW bonded Mo and W containing materials over Cr.

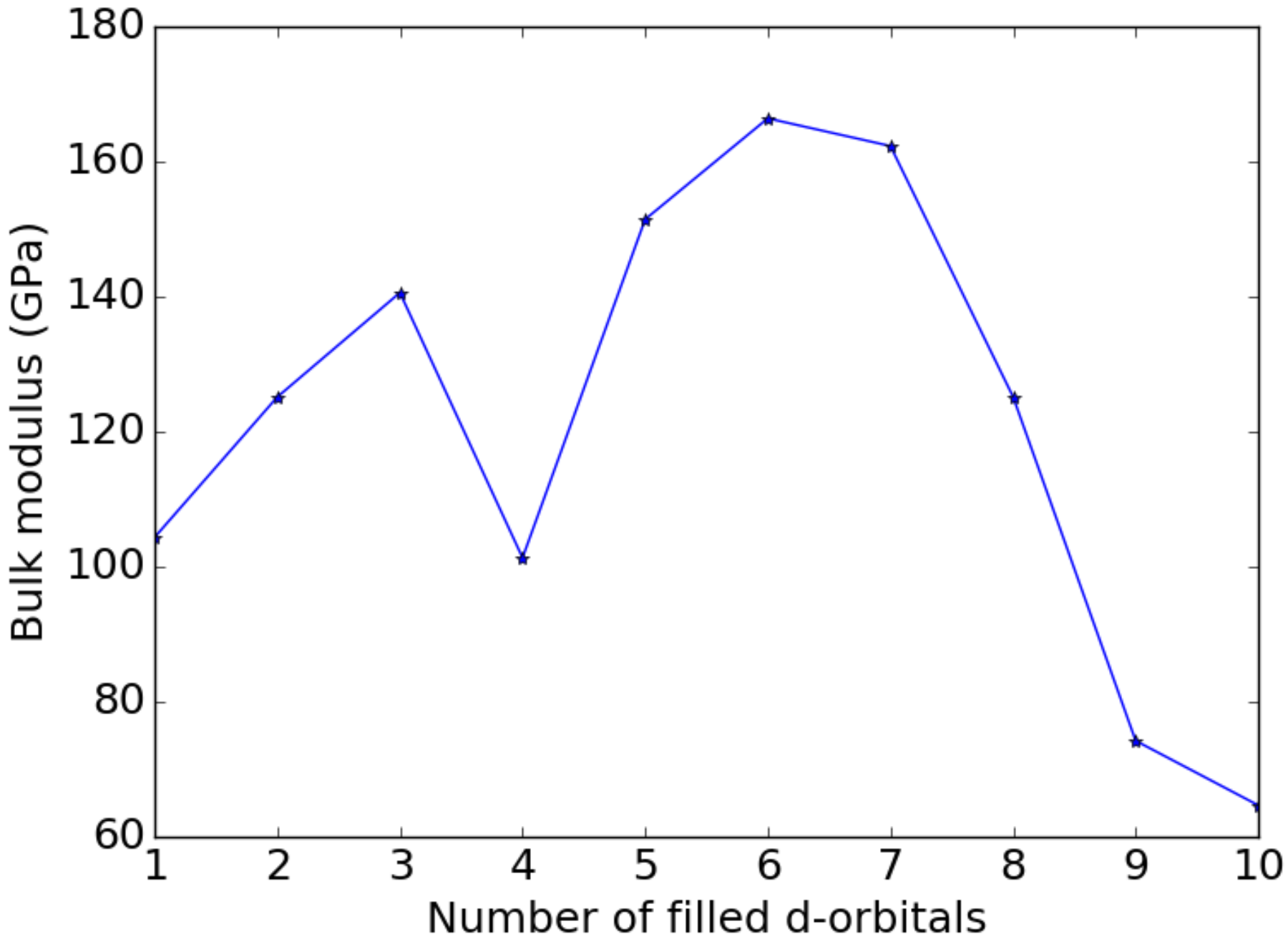

*Fig. 5 Correlation of the number of filled d-orbitals with the bulk modulus obtained by averaging the element projected bulk modulus for transition metals (shown in Fig. 4) over each periodic table column (ex: averaging the element projected bulk modulus among Ti, Zr and Hf for d=2, where d=filled d-orbitals). With the exception of W-group, the trend is very clear and is in agreement with the observed behavior of a particular group of materials (carbides, nitrides etc.)*

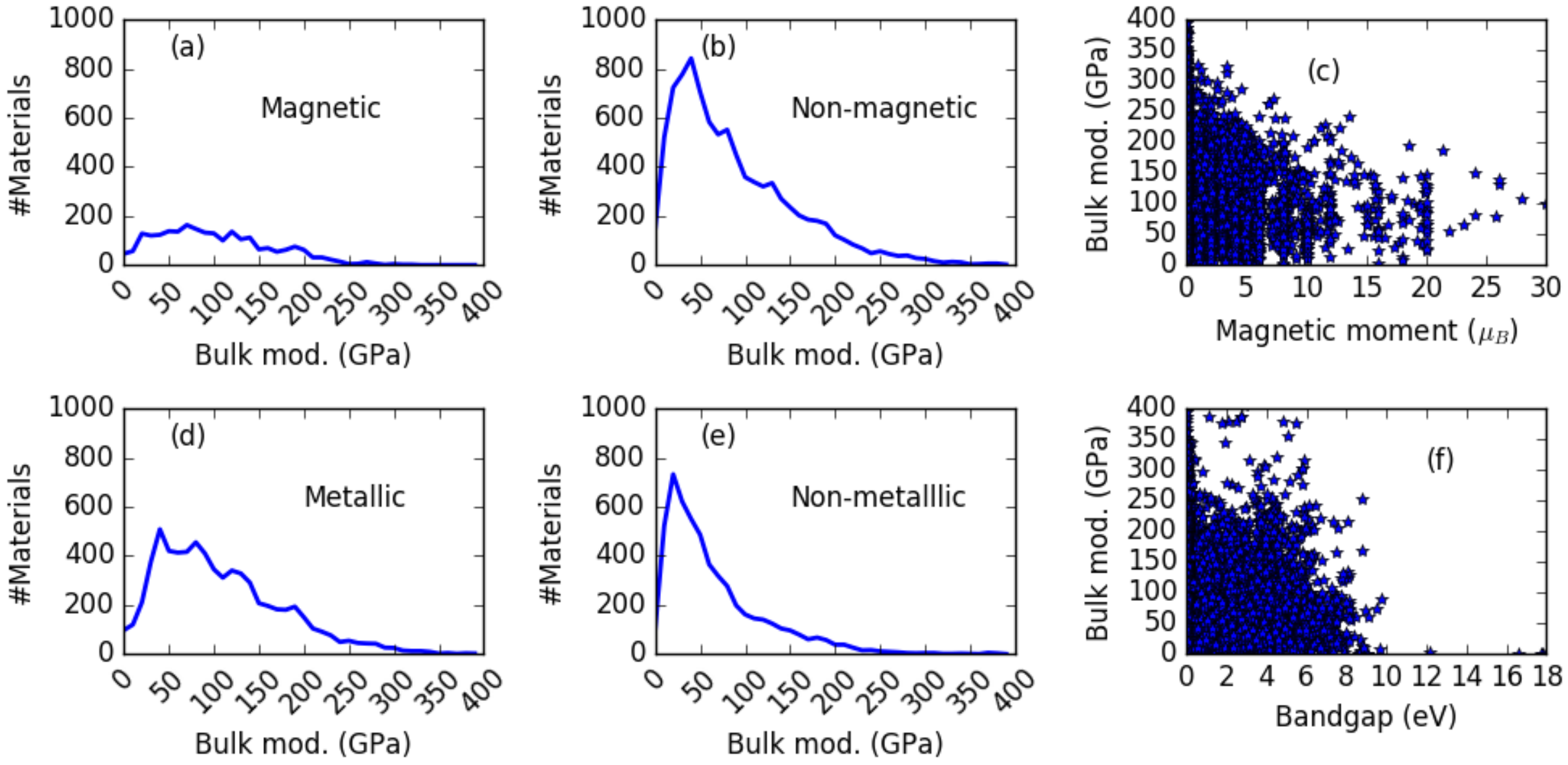

*Fig. 6 Correlation of electronic and magnetic properties (bandgap and magnetic moment) with bulk modulus.*

Next, Fig. 6 shows that the non-magnetic materials dominate the database (Fig. 6a and 6b), while the numbers of metallic (bandgap = 0, Fig. 6d) and non-metallic (bandgap >0, Fig. 6e) materials are similar. While the bulk modulus range is very similar in all cases, metallic and non-magnetic materials have relatively higher bulk modulus on average ($B_{average}$ (metals)=111 GPa, $B_{average}$ (non-metal) = 70 GPa, $B_{average}$ (magnetic)=98 GPa, $B_{average}$ (non-magnetic) = 93 GPa). We also find that the maximum bulk modulus decreases as magnetic moment (Fig. 6c) and bandgap increase (Fig. 6f). There is no clear interpretation of these trends, however, these empirical relationships can guide material discovery.

Next, we describe the elastic constant distribution of all the materials in our database for bulk 3D, 2D, 1D and 0D materials in Fig. 7. The distribution of 6x6 elastic constants for all the materials, and for 2D, 1D and 0D data is shown in magenta, green, blue and red, respectively. Firstly, we observe that nine most important elastic-constants (ECs) are $C_{11}$, $C_{22}$, $C_{33}$, $C_{12}$, $C_{13}$, $C_{21}$, $C_{31}$, $C_{44}$, $C_{55}$, and $C_{66}$, while other elastic constants seem to have very low values for distribution. Interestingly, we find that as the dimensionality decreases, the EC decreases, which can be attributed to the weak vdW bonding. The red line attains the lowest value among all the distributions implying weakest bonding in 0D materials. It is important to mention that the vdW bonding can be in *x*, *y*, *z* or any direction, however, the trend is clearly visible in the $C_{11}$, $C_{22}$, and $C_{33}$. Our individual elastic constant data can also be used to predict Born's stability for materials, elastic anisotropy, Debye temperature, the lower limit of thermal conductivity, empirical harness and Young's modulus.

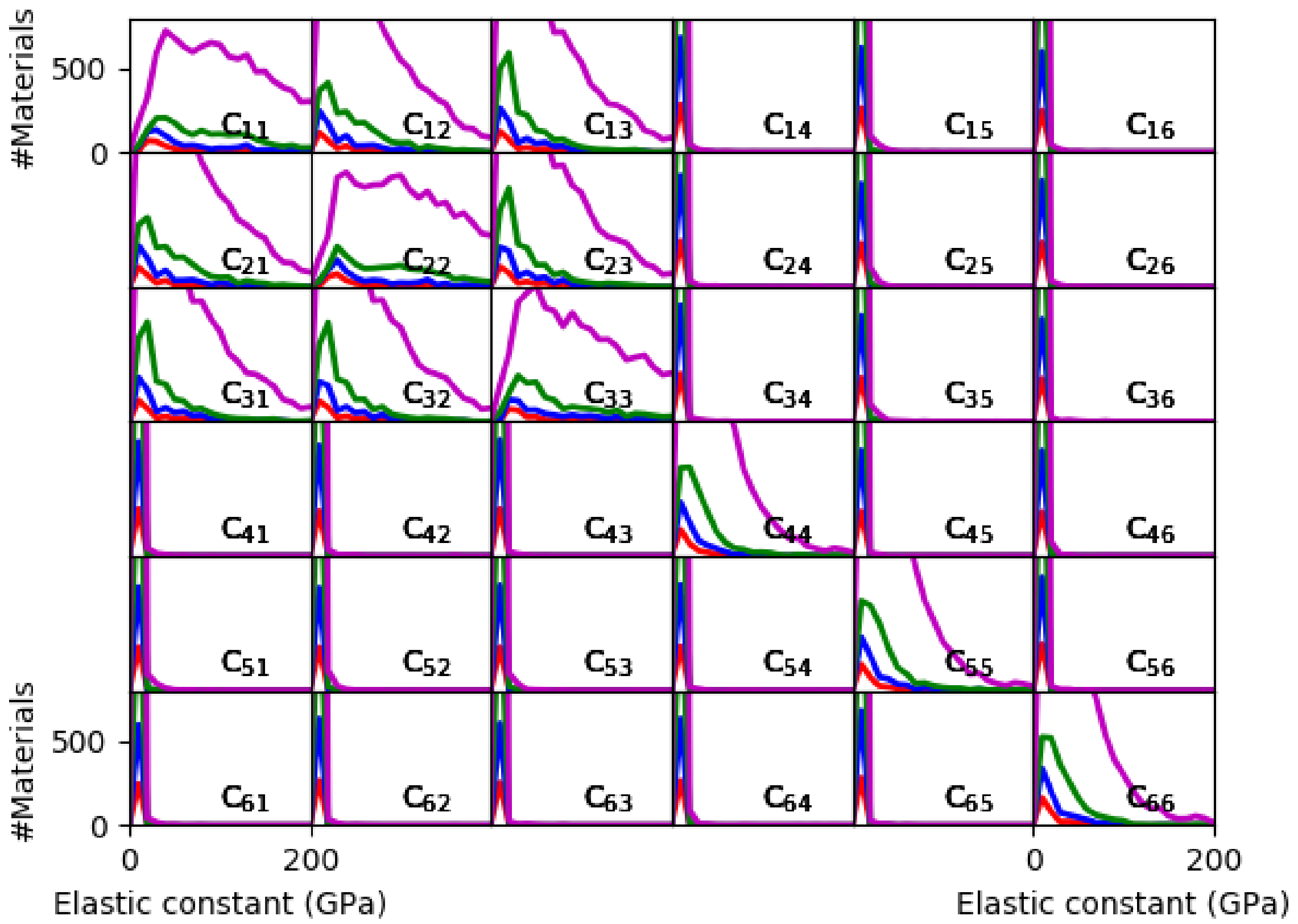

*Fig. 7 Elastic constant distribution for 3D (magenta), 2D (green), 1D (blue) and 0D (red) materials.*

In Fig. 8a, the Pugh and Pettifor criteria for all the materials are shown. We construct a convex hull based boundary regions (boundary of all the scattered points) for all the 3D, 2D, 1D and 0D materials to investigate how dimensionality of materials influences the ductile/brittle nature. Materials with Pugh ratio ($G_v/K_v$) value >0.571 and Pettifor criteria <0 should be brittle and vice-versa. Of course, data such as ultimate strength and strain are computationally expensive, these criteria can be used as a first step in the screening of materials. We clearly observe that the overall distribution of brittle and ductile materials is the same implying that our database consists of a good combination of both brittle and ductile materials. The 1D and 0D materials seem to be mainly ductile, while the 2D materials span over both the ductile and brittle regions according to the above-mentioned criteria. We explain this behavior due to the presence of weak vdW bonding which favors ductile behavior. The low dimensional materials are similar to ductile polymers [78], where vdW bonding is generally present.  In fact, some of the 2D materials exhibit ductile behavior as shown by molecular dynamics simulations [79].  In Fig. 8b, Poisson ratio distribution for all the bulk materials, 2D, 1D, and 0D are shown. Poisson ratio is a measure of compressibility of materials. As Poisson ratio approaches 0.5, the material has a tendency to become incompressible. As obvious, most of the materials are found to possess Poisson ratio between 0.1 and 0.6. However, we notice a few materials, which are predicted to have negative Poisson ratio. The negative Poisson materials are also known as auxetic materials and shown anomalous anisotropic behavior. Some of the 2D auxetic materials are also characterized recently by experiments [23] showing promising industrial applications of these materials. We predict some of new auxetic materials as: PbS (-0.5,

Cmcm, JVASP-28369), Al (-6.2, Im-3m, JVASP-25408), $CSi_2$ (-0.13, P6/mmm, JVASP-16869), $YbF_3$(-0.06, Pnma, JVASP-14313) and $SiO_2$(-0.03, $Pna2_1$, JVASP-22571). We provide the Poisson ratio, space-group information and JARVIS-ID in the parenthesis. The JARVIS-ID can be used to obtain detailed structural and electronic properties of these materials through the database. Most of these phases are not on convex hull (based on formation energy data and energy above hull data from MP), implying they might not be thermodynamically stable or high-pressure phases. Actual values of Poisson ratios are obtained through experiments, but the predicted values here can act as a guide to experiments. We also find that Poisson ratio distribution range is mostly independent of nature of dimensionality of materials as shown in Fig.7b. In addition to the homogeneous Poisson ratio discussed above, directional Poisson ratio can also be calculated from our ET data. The directional Poisson ratio for bulk materials can guide whether the materials can have negative Poisson ratios in a particular direction. For example, 2D black phosphorous has positive directional Poisson ratio in *x*-direction, but negative Poisson ratio in *y*-direction. This, in turn, can be considered as the signature of negative Poisson ratio in bulk materials that also shows up in monolayer materials such as Phosphorene [22].

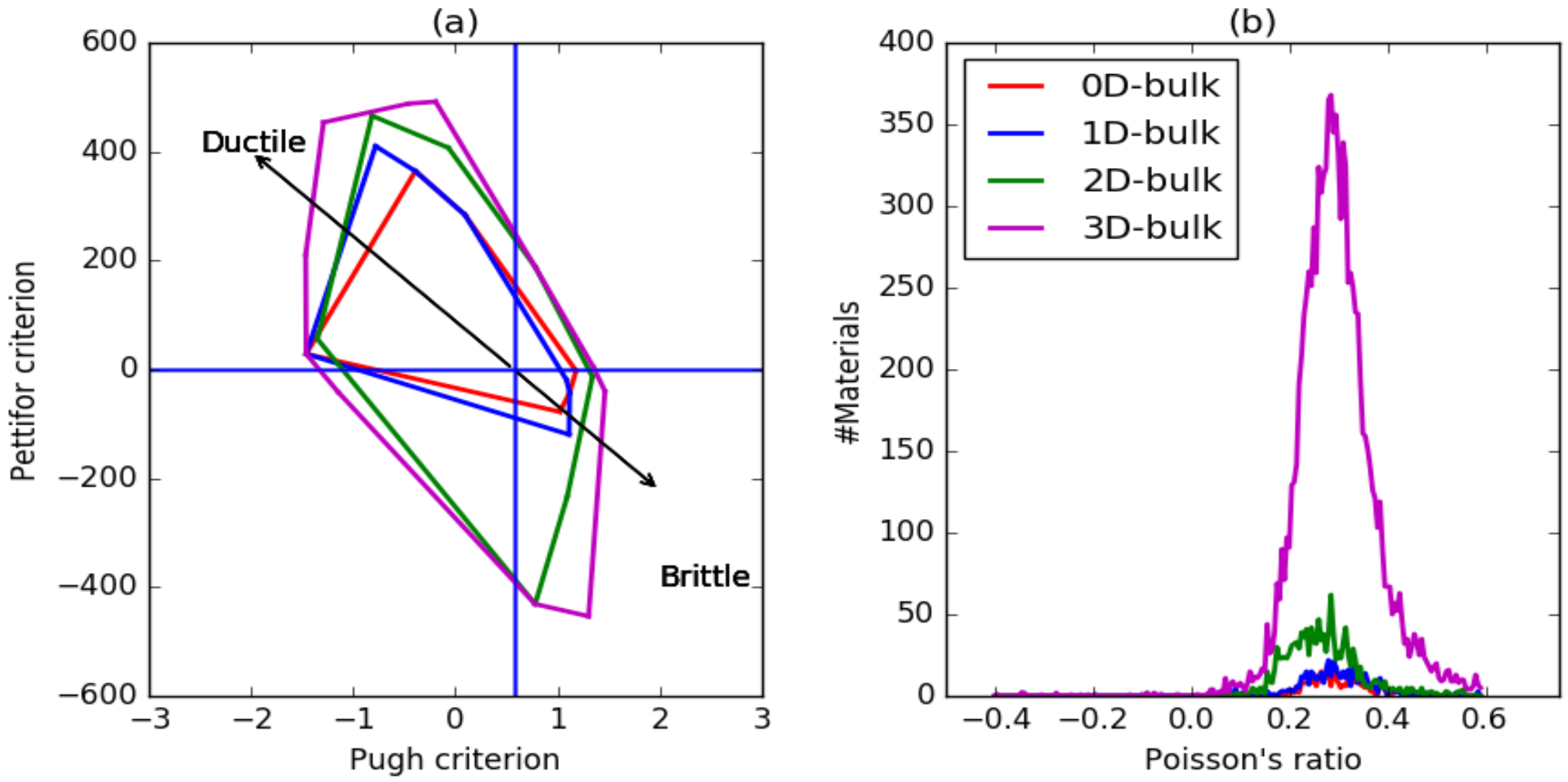

*Fig. 8 Effect of dimensionality on ductile-brittle and Poisson ratio predictions. Scatter plot boundary regions for Pugh-Pettifor criteria predicting brittle and ductile nature of materials is shown in Fig. a, while Poisson ratio distribution for 3D, 2D, 1D and 0D materials is shown in Fig. b with magenta, green, blue and red color lines respectively.*

Next, we analyze the relation between exfoliation energy obtained from our previous work [30] and presently available elastic constants. As vdW bonding can be present in any of the three crystallographic directions, we plot the minimum of elastic constants in *x*, *y*, and *z*-directions against exfoliation energy of the predicted 2D materials. The exfoliation energies were obtained by the difference in energy/atom for bulk and monolayer calculation for a particular material:

$$E_f = \frac{E_{1L}}{N_{1L}} - \frac{E_{2D-bulk}}{N_{2D-bulk}} \tag{15}$$

Here, $E_{1L}$ and $E_{2D\text{-}bulk}$ are the energies of the monolayer and 2D bulk materials and $N_{1L}$ and $N_{2D\text{-}bulk}$ are the number of the atoms in the monolayer and 2D-bulk systems respectively. As obvious from the Fig. 9, the elastic constants for the 2D materials which have exfoliation energy less than 200 meV/atom, are less than 50 GPa. This suggests low elastic constant can be considered as signatures of weak bonding such as vdW bonding in materials. In this way, low elastic constant materials can also be pre-screened as vdW materials similar to our simple lattice constant criteria and data-mining approaches mentioned above.

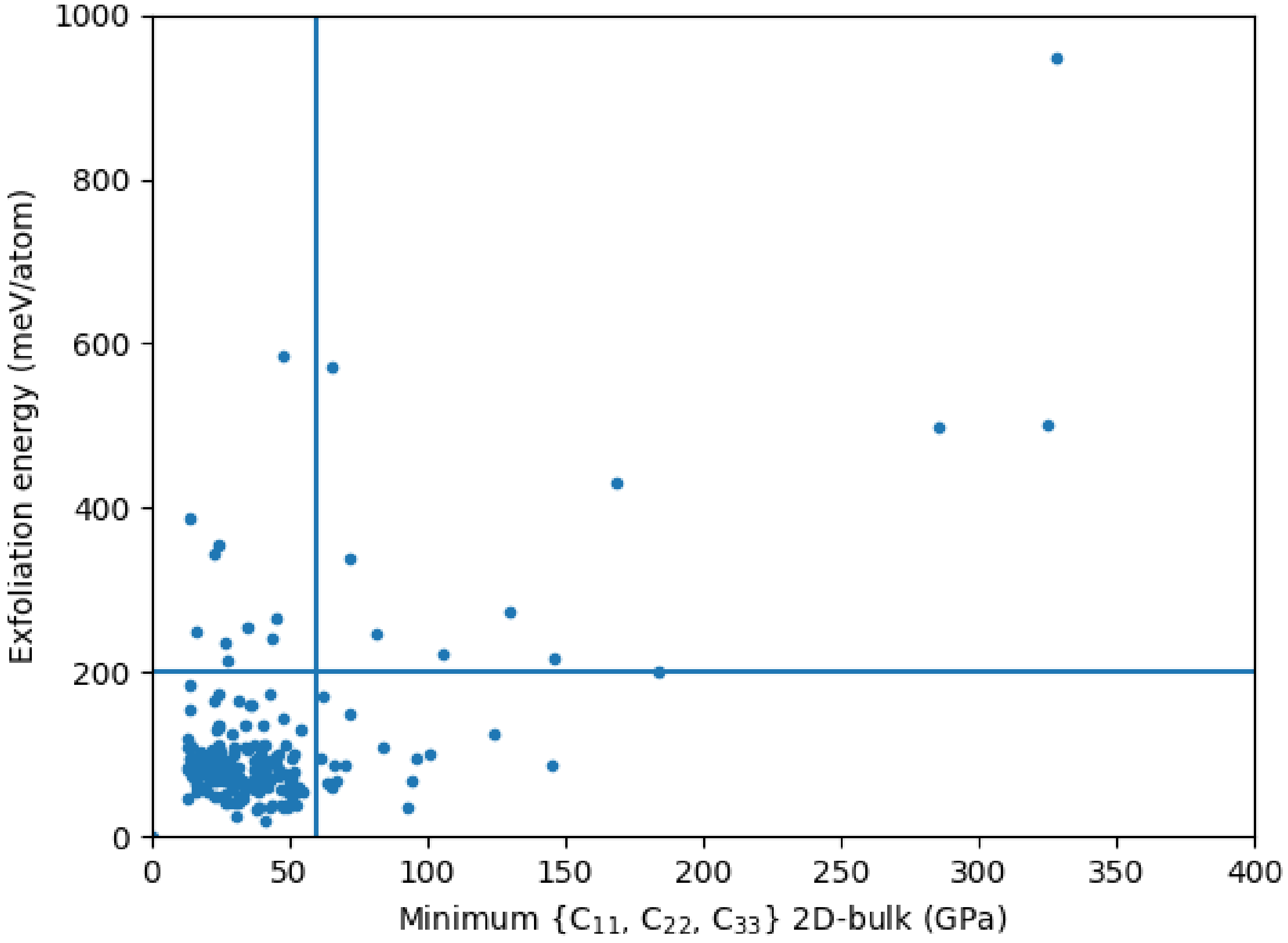

*Fig. 9 Relation of exfoliation energy with anisotropic elastic constants of bulk layered materials.*

In addition to the low-dimensional elastic constant data, we also calculate monolayer elastic constant properties of materials. It is to be noted that the elastic constants for bulk materials are volumetric quantity while that for monolayer materials, it is a surface quantity, hence expressed as $Nm^{-1}$. While computing with DFT, we give large vacuum in *z*-direction/vdW direction (>1.8 nm, enforcing z-direction to be the vdW direction) for mono- layer materials and calculate elastic tensor similar to bulk materials. However, after the calculation, we multiply the ET with the thickness of the material to get ET in $Nm^{-1}$ units for all ET components except $C_{33}$ as discussed in the method section. While the bulk and monolayer ET data may not be completely comparable, ET can be compared among the monolayer materials itself because of their consistent physical units.

Experimentally, the layer dependence of elastic constants for monolayers is compared with bulk assuming a finite thickness (such as 0.65 nm) [17,18]. In our database, we provide the elastic constant in $Nm^{-1}$ so that a user can pick arbitrary thickness to compare various bulk and monolayer materials. Experimental measurements of monolayer materials are much more challenging than their bulk counterparts, hence, there are only a few such data available right now. Some of the experimental measurements for $C_{11}$ include: graphene [13] (340 $Nm^{-1}$), $MoS_2$ [17,80] (180 $\pm$ 60 $Nm^{-1}$and 130 $Nm^{-1}$), $WS_2$ [81] (177 $\pm$ 12 $Nm^{-1}$) and BN [82] (289±24 $Nm^{-1}$) . Our DFT results for these materials are: 354.6 $Nm^{-1}$ for graphene (JVASP-667), 134.3 $Nm^{-1}$ for $MoS_2$ (JVASP-664), 146.5 $Nm^{-1}$ for $WS_2$ (JVASP-658) and 293.2 $Nm^{-1}$ for BN (JVASP-688) showing an excellent agreement between our DFT data and experiments.

The $C_{11}$ and $C_{12}$ values are generally the most important elastic constants for monolayer materials [24]. Therefore, we provide a distribution of $C_{11}$ and $C_{12}$ for monolayer materials in Fig. 10. We observe that most of the $C_{11}$ for monolayer materials are around 100 $Nm^{-1}$ but it can be as high as 400 $Nm^{-1}$. The $C_{12}$ has more localized distribution than $C_{11}$. Some of the high $C_{11}$ monolayer materials are C (354.9 $Nm^{-1}$, JVASP-667), BN (293.3 $Nm^{-1}$, JVASP-688), $Ta_2Se$ (219.3 $Nm^{-1}$, JVASP-13541), $NbIO_2$ (181.8 $Nm^{-1}$, JVASP-28028), HfIn (176.4 $Nm^{-1}$, JVASP-27774), $Si_3H$ (169.8 $Nm^{-1}$, JVASP-14451) and HfNCl (166 $Nm^{-1}$, JVASP-13477), AlClO (161.5 $Nm^{-1}$, JVASP-6271). Some of the low $C_{11}$ materials are: AgI (18.1 $Nm^{-1}$, JVASP-14417), InBi (20 $Nm^{-1}$, JVASP-31353), AuBr (21.0 $Nm^{-1}$, JVASP-27756), $VBr_2$ (23.25 $Nm^{-1}$, JVASP-13546), Bi (25.28 $Nm^{-1}$, JVASP-20002), $CdCl_2$ (30.0 $Nm^{-1}$, JVASP-6232). The JARVIS-ID in the parenthesis can be used to obtain detailed structural and electronic properties of these materials through our database.

All the monolayer calculation data are available on our website, and the database is still developing with the promise to contain elastic constants of thousands of such layered materials. While high elastic constant monolayer-materials can be used for designing stiff materials, low elastic constant materials can be used for flexible materials applications [83]. The strength of some materials such as graphene decreases dramatically with an increase in thickness, but few-layer BN nanosheets (at least up to 9L) have a strength similar to that of 1L BN [82]. Therefore, understanding how ET changes as the number of layers changes does is an interesting issue and will be investigated in the future.

As we use the finite-difference method to calculate elastic constant, all the finite-size gamma-point phonon data obtained during the calculations are also reported on the website. Phonons with highly negative frequencies indicate the dynamic instability of materials, hence, we provide all such data on webpages for each material. In addition, the convex hull stability of materials can be used to investigate the thermodynamic stability of materials. At present, we have not provided the convex hull energy values for all the materials, but the formation energies of materials available on our website can be used to compute convex-hull stability. Moreover, a user can also use our 6x6 elastic constants data to predict Born's elastic constant stability [8] of materials.

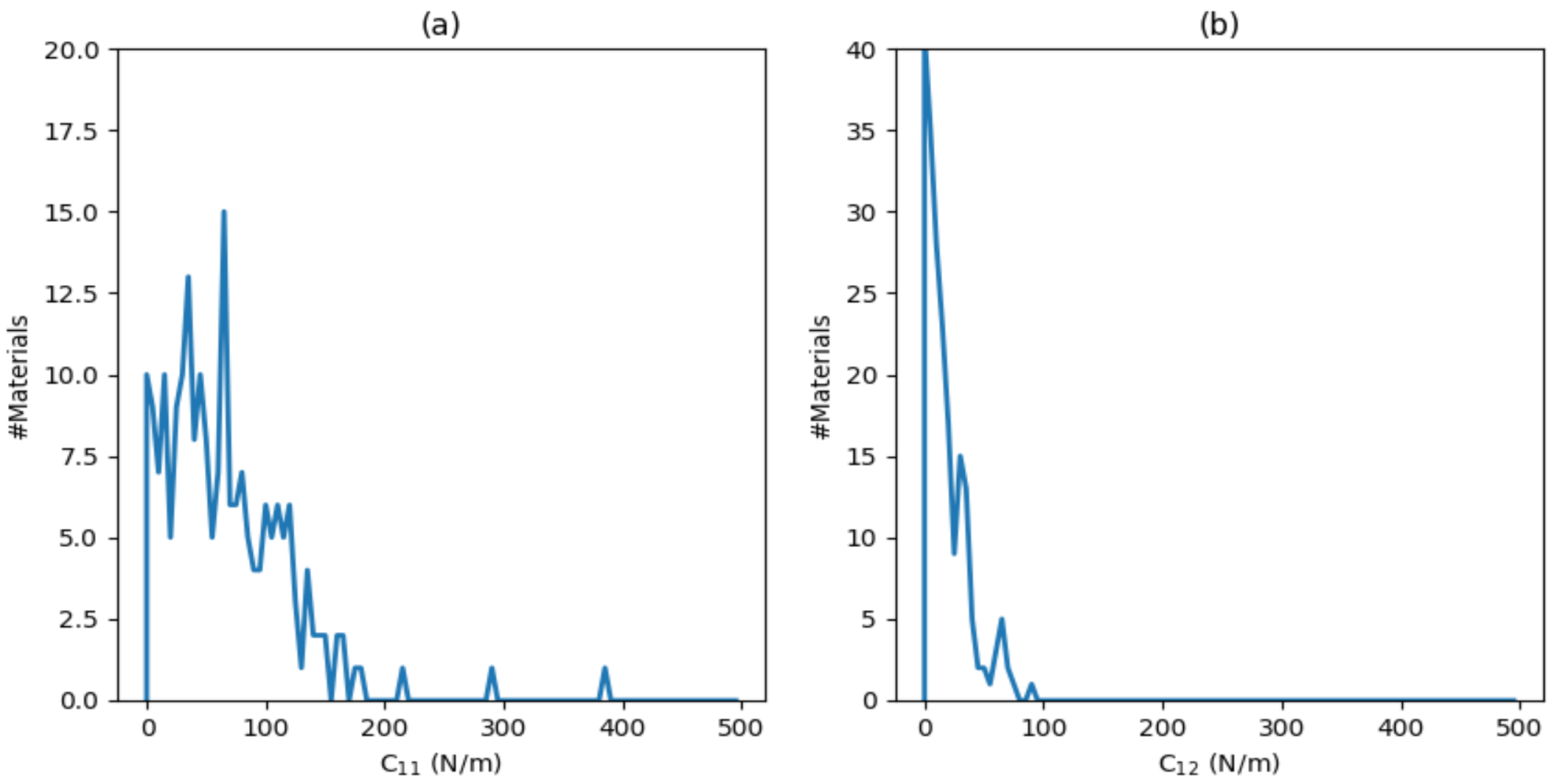

*Fig. 10 Elastic constant distributions ($C_{11}$ and $C_{12}$) for monolayer (1L) materials.*

Next, we investigate if the ranking order of materials remains the same as we create monolayers from their bulk counterparts. We sorted the bulk and corresponding 1L elastic constants and show some of them in Table. 3 to find the trends. From Table. 3 we observe that the monolayer elastic constants ranking can change drastically compared to their bulk counterparts. It also shows that the elastic response changes as we exfoliate a vdW bonded material. Our data can also be used to understand mismatch in heterostructures [10]. Previously, Gomes et al. [25] established the comparison of bulk to monolayer elastic constant should be done by dividing the $C_{11}$ of monolayers by layer thickness. However, the layer thickness can be a complex issue for materials other than simple monolayer materials such as graphene [13]. Much work still needs to be done in standardizing the comparison of layer-dependent and bulk material data. Hence, we provide the raw data to users to facilitate their own comparison.

*Table. 3 Order comparison for $C_{11}$ of bulk and their monolayer counterpart for a few materials in our database.*

| Materials | 1L-$C_{11}$ ($Nm^{-1}$) | 3D-bulk-$C_{11}$ (GPa) | JARVIS-IDs |
|---|---|---|---|
| **C** | 354.85 | 1058.9 | JVASP-667, JVASP-48 |
| **BN** | 293.25 | 883.7 | JVASP-688, JVASP-17 |
| **$Ta_2Se$** | 219.34 | 228.5 | JVASP-13541, JVASP-12179 |
| **$NbIO_2$** | 181.94 | 180.1 | JVASP-28028, JVASP-25591 |
| **HfIN** | 176.41 | 170.4 | JVASP-27774, JVASP-12131 |
| **$Si_3H$** | 169.65 | 149.8 | JVASP-14451, JVASP-12058 |
| **$AlHO_2$** | 161.7 | 269.4 | JVASP-14432, JVASP-12038 |
| **AlClO** | 161.63 | 207.0 | JVASP-6271, JVASP-13787 |
| **ZrNCl** | 151.95 | 169.5 | JVASP-27777, JVASP-12136 |
| **$Sc_2CCl_2$** | 150.16 | 171.5 | JVASP-6172, JVASP-3993 |
| **$WS_2$** | 146.48 | 233.3 | JVASP-658, JVASP-72 |

**Conclusions:**

We evaluated the trends in elastic properties and derived quantitates for three-dimensional as well as monolayer materials using the vdW-DF-optB88 (OPT) functional. Low-dimensional materials are found to have a decreasing order of elastic constants with respect to a decrease in dimensionality. The trends in elastic properties in presence of vdW bonding in multiple directions are discussed and can be used in designing high/low strength materials. We predicted a few novel materials that have auxetic behavior. We also establish the relation between elastic constants and exfoliation energies of 2D-bulk materials. At present, we have 11067 bulk and 257 monolayer elastic constant data. We find that the order of elastic constants for bulk and their single-layer counterparts can be very different implying the importance of single layer elastic constants. Our database is publicly available on the websites: https://jarvis.nist.gov and https://www.ctcms.nist.gov/~knc6/JVASP.html. Data mining, data analytics, and machine learning tools can further be applied to guide screening of materials.

# Supplementary information: Elastic properties of bulk and low-dimensional materials using Van der Waals density functional

Kamal Choudhary[1], Gowoon Cheon[2], Evan Reed[2], Francesca Tavazza[1]

1 Materials Science and Engineering Division, National Institute of Standards and Technology, Gaithersburg, Maryland 20899, USA

2 Department of Materials Science and Engineering, Stanford University, Stanford, California 94305, United States

Table S1 Bulk modulus of materials with several functionals in vdW-DF [34,52-59]: vdW-DF-optB88 (OPT), vdW-DF-optB86b (MK), vdW-DF-optPBE (OR) and vdW-DF-cx13 (CX). Highest mean absolute error (MAE) error is obtained for OR. The OPT and CX gives very similar MAEs. It is to be noted that MK was shown to have a slightly high error in binding energies [55].

| **Materials** | **OPT** | **MK** | **OR** | **CX** | **Expt.** |
|---|---|---|---|---|---|
| Cu | 141.4 | 158.2 | 139.3 | 166.7 | 142 |
| C | 437.4 | 435.5 | 422.2 | 439.0 | 443 |
| Si | 87.3 | 91.1 | 86.9 | 92.8 | 99.2 |
| Ge | 58.1 | 63.1 | 56.4 | 66.4 | 75.8 |
| Ag | 100.3 | 107.4 | 87.3 | 116.3 | 109 |
| Pd | 176.0 | 187.8 | 161.5 | 200.1 | 195 |
| Rh | 260.8 | 277.8 | 248.9 | 291.7 | 269 |
| Li | 13.9 | 13.6 | 14.4 | 13.3 | 13.3 |
| K | 3.9 | 3.8 | 3.8 | 3.2 | 3.7 |
| Rb | 3.1 | 3.1 | 3.1 | 2.6 | 2.9 |
| Ca | 17.7 | 17.9 | 17.4 | 17.4 | 18.4 |
| Sr | 12.5 | 12.4 | 12.0 | 12.1 | 12.4 |
| Ba | 9.9 | 9.6 | 9.4 | 9.4 | 9.3 |
| Al | 70.0 | 77.1 | 71.2 | 79.6 | 79.4 |
| LiF | 73.9 | 71.9 | 70.4 | 74.8 | 69.8 |
| LiCl | 35.5 | 34.8 | 33.4 | 34.9 | 35.4 |
| NaCl | 27.7 | 26.9 | 26.2 | 24.6 | 26.6 |
| NaF | 53.7 | 50.9 | 49.9 | 45.0 | 51.4 |
| MgO | 160.7 | 158.0 | 153.2 | 153.8 | 165 |
| SiC | 213.3 | 215.5 | 208.2 | 217.9 | 225 |
| GaAs | 62.0 | 63.5 | 57.6 | 66.9 | 75.6 |
| MAE | 5.75 | 4.66 | 9.18 | 5.85 | - |

Table S2 Comparison of 0K $C_{11}$ (GPa) for OPT and CX compared to experiments at low-temperature (4 to 6 K) and room-temperature (300K). The MAE of OPT is similar to that of CX (calculated with respect to low temperature experimental data). DFT values are generally higher than experimental data at low temperatures. Elastic constants at low temperatures are generally higher than that at room temperatures. Low-temperature data were taken from ref. [84] and room-temperature data from ref. [85-93].

| **Materials** | **$C_{11}$-OPT** | **$C_{11}$-CX** | **Expt-6K** | **Expt-300K** |
|---|---|---|---|---|
| Al | 93.4 | 99.9 | 122.96 | 107.3 |
| Cu | 175.8 | 207.9 | 176.2 | 171.2 |
| K | 4.1 | 4.06 | 4.16 | 3.70 |
| LiCl | 62.0 | 65.2 | 60.74 | 48.3 |
| LiF | 129.6 | 138.2 | 124.45 | 111.2 |
| NaCl | 61.0 | 54.9 | 58.38 | 49.8 |
| NaF | 117.3 | 99.3 | 110.39 | 97.0 |
| Pd | 208.5 | 240.4 | 234.12 | 223.8 |
| Rb | 3.3 | 2.75 | 3.42 | 2.41 |
| MAE (wrt 6K) | 7.97 | 10.51 | - | 8.9 |
| MAE (wrt 300K) | 10.9 | 12.5 | 8.9 | - |

Table S3. Elastic constant comparison between DFT OPT data and experimental measurements. Experimental measurement data were taken from [94,95]

| **Mats.** | **JVASP** | **SG** | **$C_{11}$-DFT** | **$C_{11}$-Exp** | **$C_{12}$-DFT** | **$C_{12}$-Exp** | **$C_{33}$-DFT** | **$C_{33}$-Exp** | **$C_{44}$-DFT** | **$C_{44}$-Exp** |
|---|---|---|---|---|---|---|---|---|---|---|
| C | 91 | Fd-3m | 1061 | 1079 | 125.6 | 124 | 1061 | 1079 | 559.3 | 578 |
| Si | 1002 | Fd-3m | 150 | 168 | 56 | 65 | 150 | 168 | 74 | 80 |
| Ge | 890 | Fd-3m | 100.6 | 124 | 36.8 | 41.3 | 100.6 | 124 | 52.0 | 68.3 |
| Sn | 1008 | Fd-3m | 54.1 | 69 | 27.7 | 29 | 54.1 | 69 | 24.1 | 36 |
| AlAs | 1372 | F-43m | 105.4 | 120.2 | 50.6 | 57 | 105.4 | 120.2 | 52.7 | 58.9 |
| AlSb | 1408 | F-43m | 76.9 | 89.4 | 37.2 | 44.3 | 76.9 | 89.4 | 36.8 | 41.6 |
| GaAs | 1174 | F-43m | 100 | 119 | 42.9 | 53.8 | 100 | 119 | 51.5 | 59.5 |
| InAs | 97 | F-43m | 71.1 | 83.3 | 39.5 | 45.3 | 71.1 | 83.3 | 32.8 | 39.6 |
| InSb | 1189 | F-43m | 54.6 | 66.9 | 29.9 | 36.7 | 54.6 | 66.9 | 24.7 | 30.2 |
| ZnO | 1195 | $P6_3mc$ | 195.2 | 207 | 116.1 | 117.7 | 211.8 | 209.5 | 39.5 | - |
| ZnS | 7648 | $P6_3mc$ | 118.3 | 123.4 | 55.5 | 58.5 | 28.85 | 28.9 | 31.4 | 32.45 |
| CdSe | 7671 | $P6_3mc$ | 67.1 | 74.0 | 41.6 | 45.2 | 77.7 | 83.6 | 12.7 | 13.2 |
| CdTe | 7757 | $P6_3mc$ | 54.7 | 53.3 | 31.9 | 36.5 | 65.7 | - | 11.4 | 19.4 |
| $MoS_2$ | 54 | $P6_3/mmc$ | 214.6 | 238 | 57.2 | -54 | 48.2 | 52 | 78.7 | 19 |
| MAE | | | 15.04 | - | 14.27 | - | 14.44 | - | 12.12 | - |

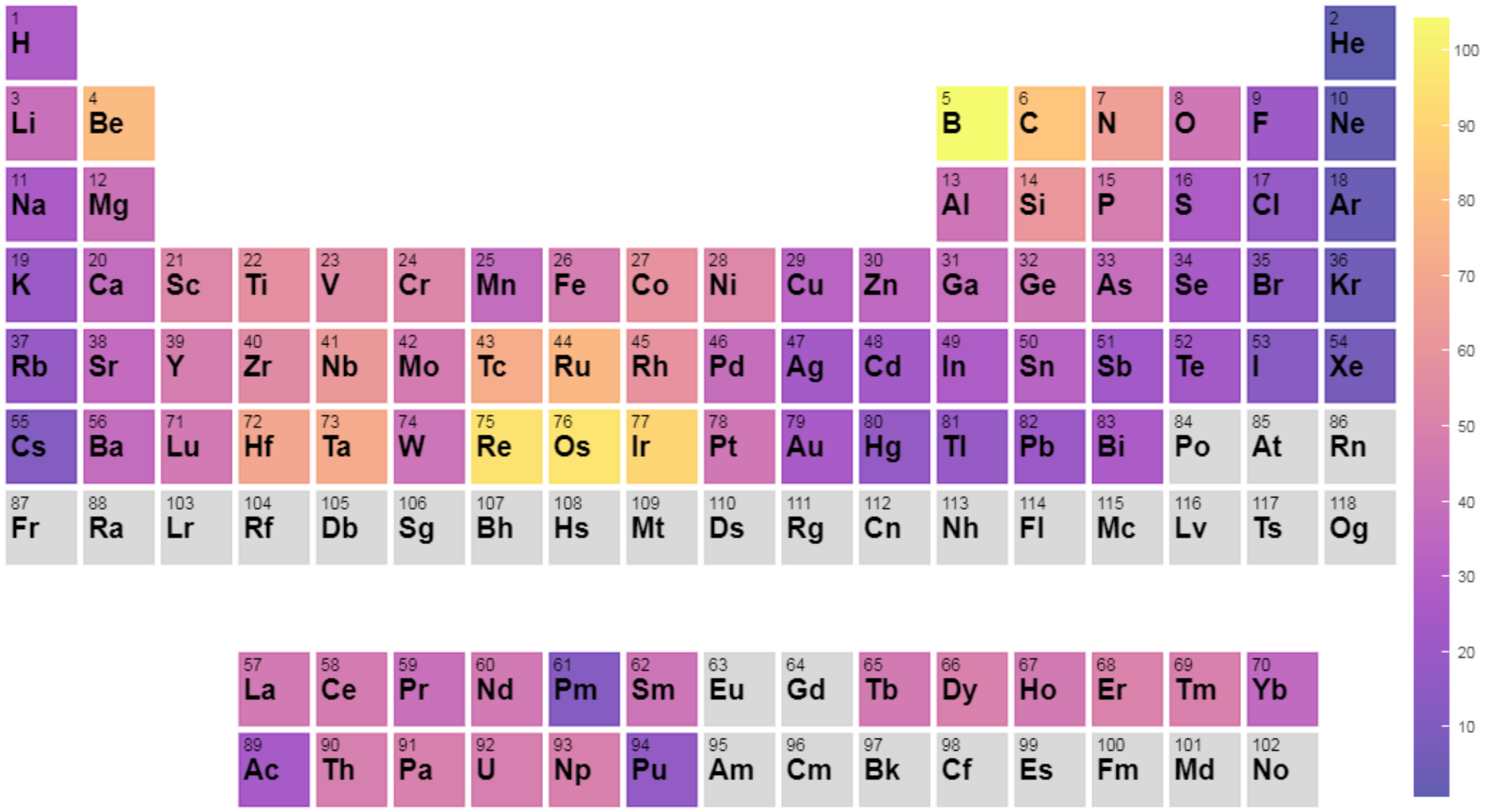

Fig. S1 Periodic table trend for shear modulus material constituents. Shear modulus of all the materials were projected on individual elements and their average contribution is shown. The colorbar is in the unit of GPa.